\documentclass[aps,preprint,superscriptaddress,showpacs,nofootinbib]{revtex4-1}
\usepackage{graphicx}
\usepackage{amsfonts}
\usepackage{mathrsfs}
\usepackage{epsfig}
\usepackage{slashed}
\usepackage{dcolumn}%

\usepackage{ulem}
\usepackage{color}
\usepackage{caption}
\usepackage{subcaption}
\usepackage[colorlinks,
linkcolor=blue,
anchorcolor=blue,
citecolor=green
]{hyperref}
\definecolor{My_red}        {cmyk}{0.00,1.00,1.00,0.20}

\newcommand{\bmat}{\left(\begin{array}}
\newcommand{\emat}{\end{array}\right)}
\newcommand{\beq}{\begin{equation}}
\newcommand{\eeq}{\end{equation}}

\def\bwt{\begin{widetext}}
\def\ewt{\end{widetext}}
\def\be{\begin{equation}}
\def\ee{\end{equation}}
\def\bea{\begin{eqnarray}}
\def\eea{\end{eqnarray}}
\def\bean{\begin{eqnarray*}}
\def\eean{\end{eqnarray*}}
\def\bary{\begin{array}}
\def\eary{\end{array}}
\def\bit{\begin{itemize}}
\def\eit{\end{itemize}}

\def\su5u1{SU(5) \times U(1)}
\def\fsu5u1{SU(5) \times U(1)'}
\def\so10{SO(10)}
\def\sq20{SO(10) \times SO(10)}

\def\bwt{\begin{widetext}}
\def\ewt{\end{widetext}}
\def\be{\begin{equation}}
\def\ee{\end{equation}}
\def\bea{\begin{eqnarray}}
\def\eea{\end{eqnarray}}
\def\bean{\begin{eqnarray*}}
\def\eean{\end{eqnarray*}}
\def\bary{\begin{array}}
\def\eary{\end{array}}
\def\bit{\begin{itemize}}
\def\eit{\end{itemize}}

\def\su5u1{SU(5) \times U(1)}
\def\fsu5u1{SU(5) \times U(1)'}
\def\so10{SO(10)}
\def\sq20{SO(10) \times SO(10)}

\usepackage{amsmath}
\usepackage{amssymb}
\usepackage{color}

\begin{document}

\title{The generic $U(1)_X$ models inspired from $SO(10)$}

\author{Tianjun Li}
\email{tli@itp.ac.cn}
\affiliation{
CAS Key Laboratory of Theoretical Physics, Institute of Theoretical Physics,
Chinese Academy of Sciences, Beijing 100190, China
}
\affiliation{
School of Physical Sciences, University of Chinese Academy of Sciences,
No.~19A Yuquan Road, Beijing 100049, China
}
\author{Qianfei Xiang}
\email{xiangqf@pku.edu.cn}
\affiliation{
CAS Key Laboratory of Theoretical Physics, Institute of Theoretical Physics,
Chinese Academy of Sciences, Beijing 100190, China
}

\author{Xiangwei Yin}
\email{yinxiangwei@itp.ac.cn}
\affiliation{
CAS Key Laboratory of Theoretical Physics, Institute of Theoretical Physics,
Chinese Academy of Sciences, Beijing 100190, China
}
\affiliation{
School of Physical Sciences, University of Chinese Academy of Sciences,
No.~19A Yuquan Road, Beijing 100049, China
}

\author{Han Zhou}
\email{zhouhan@alumni.itp.ac.cn}
\affiliation{
CAS Key Laboratory of Theoretical Physics, Institute of Theoretical Physics,
Chinese Academy of Sciences, Beijing 100190, China
}


\begin{abstract}

We propose the family universal $U(1)_X$ models with three right-handed neutrinos
by choosing the $U(1)_X$ gauge symmetry as a linear combination of $U(1)_Y\times U(1)_{\chi}$ 
of $SO(10)$. To be consistent with the quantum gravity effects, 
we introduced a Dirac fermion  $\chi$ as a dark matter candidate,
which is odd under the gauged $Z_2$ symmetry after $U(1)_X$ breaking.
The isospin violation dark matter with $f_n/f_p = -0.7$ can be realized naturally,
and thus the LUX, PANDAX, and XENON1T experimental constraints can be evaded.
Moreover, we study the masses and mixings for Higgs and gauge bosons, 
consider the LHC constraints on the $Z'$ mass, simulate various constraints 
from dark matter direct and indirect detection experiments, and then 
present the viable parameter spaces.
To study the LHC $Z'$ mass bounds on the generic $U(1)_X$ models, we considered four kinds of scenarios,
where scenario II with  zero  $U(1)_X$ charge for right-handed up-type quarks 
can relax the LHC $Z'$ mass bound a little bit.

\end{abstract}


\preprint{ OSU-HEP-18-nn}

\maketitle

\section{Introduction}

The Standard Model (SM) has been confirmed since Higgs particle was discovered at
the LHC. However, there exist some evidences for new physics beyond the SM, 
for example, neutrino masses and mixings, dark matter (DM), dark energy, and cosmic 
inflation, etc. Therefore, the SM is not the whole story, and 
we need to explore the new physics. There are many possible directions to 
go beyond the SM. For example, the fine-tuning problem such as gauge hierarchy problem 
leads to supersymmetry~\cite{Dimopoulos:1981au}, 
technicolor~\cite{Weinberg:1975gm}, extra dimensions~\cite{ArkaniHamed:1998rs, Randall:1999ee}, etc,
while the aesthetic issues such as the unification of fundamental interactions 
and the explanation of charge quantization
lead to the grand unified theories (GUTs)~\cite{Pati:1974yy} and string theory~\cite{GSW, JP}.

 On the other hand,
DM particle candidates have a very wide mass range from around $10^{-22}$ eV to 
about $10M_{\odot}$ mass~\cite{Lin:2019uvt}, including 
the weakly interacting massive particle (WIMP), the lightest supersymmetric particle (LSP), 
massive compact halo object (MACHO), superheavy candidates, axino, sterile neutrino, 
fuzzy DM, and etc. Among these huge amount of DM candidates, 
WIMP is a well-motivated DM candidate. It is stable, nonrelativistic, electrically neutral, 
colorless, and have a mass range from about $10$ GeV to few TeV. However,
if the discrete symmetry, which stabilizes the DM candiate, is not a gauged discrete symmetry,
it can be broken via the non-renormalizable operators due to quantum gravity effects,
and then the DM candidate can decay and cannot be a valid DM candidate.
Moreover, there are strong constraints from direct search experiments, for instance, 
the PandaX-II (2017)~\cite{Cui:2017nnn}, LUX (2017)~\cite{Akerib:2016vxi}, 
and XENON1T (2018)~\cite{Aprile:2018dbl} experiments. As we know,
the isospin-violating dark matter (IVDM) is a kind of DM with different couplings $f_p$ and $f_n$
respectively to proton and neutron, and was originally proposed to explain the tensions 
among DAMA/LIBRA, CoGeNT, and XENON experiments for light DM~\cite{Kurylov:2003ra,Giuliani:2005my}. 
Interestingly, it can evade the LUX, PANDAX, and XENON1T experimental constraints
naturally as well~\cite{Chang:2010yk,Kang:2010mh,Feng:2011vu}. 
For these xenon based experiments~\cite{Cui:2017nnn,Akerib:2016vxi,Aprile:2018dbl}, 
the ratio $f_n/f_p$ is about $-0.7$. Moreover, 
several IVDM models have been proposed in recent years~\cite{Hamaguchi:2014pja,Drozd:2015gda,Frandsen:2011cg,Belanger:2013tla,Martin-Lozano:2015vva,Kang:2010mh,Gao:2011ka,Crivellin:2015bva, Li:2019sty}.
However, it is well known that in the $U(1)'$ models with $E_6$ origin, the vector coupling of the
up-type quarks to the $Z'$ boson should be zero while their axial coupling may have nonzero value. 
Thus, one cannot realize the isospin violation with $f_n/f_p = -0.7$ in the $U(1)'$ model from $E_6$
unless one introduces vectorlike particles~\cite{Li:2019sty}.

In this paper, to explore the new physics beyond the SM, we choose a conservative approach by
neglecting the fine-tuning and aesthetic problems, and concentrate on the low energy
new physics~\cite{Davoudiasl:2004be}. In particular, we consider the family universal $U(1)_X$ models.
If we only have SM fermions, the only $U(1)_X$ model, 
which one can build, is the top hypercharge model or its variation~\cite{Chiang:2007sf, Allanach:2018lvl}.
Thus, we study the family universal $U(1)_X$ models with three right-handed
neutrinos in general, and then the neutrino masses and mixings can be explained 
via type I seesaw mechanism~\cite{Minkowski:1977sc, seesaw}. As we know, one family of the SM fermions plus
the right-handed neutrino forms a spinor representation ${\bf 16}$ in $SO(10)$ model,
 and $SO(10)$ has a subgroup $SU(3)_C\times SU(2)_L\times U(1)_Y\times U(1)_{\chi}$.
Therefore, we construct this kind of the $U(1)_X$ models by choosing the $U(1)_X$ gauge symmetry
as a linear combination of $U(1)_Y\times U(1)_{\chi}$. For more discussions about generic and particular $U(1)$ models see~\cite{Appelquist:2002mw}.
Also, we introduce another Dirac fermion  $\chi$ as a DM candidate,
which is odd under the gauged $Z_2$ symmetry after $U(1)_X$ breaking. Thus,
 $\chi$ is a viable DM candidate consistent with the quantum gravity effects.
Interestingly,
we show that the isospin violation DM with $f_n/f_p = -0.7$ can be realized naturally
in our models without introducing any vectorlike particles. Moreover,
 we study the masses and mixings for Higgs and gauge bosons, 
 consider the LHC constraints on the $Z'$ mass, 
simulate various constraints from DM direct and indirect detection experiments, 
and present the viable parameter spaces. 
To study the LHC $Z'$ mass bounds on the generic $U(1)_X$ models, we consider four kinds of scenarios:
scenarios I, II, and III have zero  $U(1)_X$ charges, respectively,
for quark doublets, right-handed up-type quarks,
and right-handed down-type quarks, and scenario IV has approximately equal charges for all the quarks.
We find that the low bounds on the $Z'$ masses are about 4.94, 4.87, 5.34, and 5.09 TeV 
for scenarios I, II, III, and IV, respectively. Thus, scenario II 
can relax the LHC $Z'$ mass bound a little bit, but not too much.

The this paper is organized as follows.
In Sec~\ref{sec:model2}, we describe the $SO(10)$-inspired $U(1)_X$ model, 
and calculate the Higgs mass and other paremeters. In Sec~\ref{sec:exp}, we study 
the constraints from the LHC, dark matter direct and indirect detection experiments 
by considering isospin violation effects. In Sec~\ref{Neutrino}, we discuss the LHC bounds on the $Z'$ masses in four kind of scenarios.
Finally, we conclude in Sec~\ref{sec:summary}.

\section{The Generic $U(1)_X$ Models Inspired from $SO(10)$} \label{sec:model2}

\begin{table}[t]
\begin{center}
\begin{tabular}{|c|c|c|c|}
\hline  SM particles & $2 \sqrt{10} Q_{\chi}$ & SM particles & $2 \sqrt{10} Q_{\chi}$ 

\\
\hline
   $Q_i, U_i^c, E_i^c$ & --1   &
            $D_i^c, L_i$  & 3    \\
\hline
           $N_i^c$             & --5   &
$H$ & --2 \\
\hline
\end{tabular}
\end{center}
\caption{The  $U(1)_{\chi}$ charges of the SM particles.}
\label{SO10Qcharge}
\end{table}

First, let us explain the convention. We denote the left-handed quark doublets,
right-handed up-type quarks, right-handed down-type quarks, 
left-handed lepton doublets,
right-handed neutrinos, right-handed charged leptons, and Higgs particle as
$Q_i^c$, $U_i^c$, $D_i^c$, $L_i$, $N_i^c$, $E_i^c$, and $H$, respectively.
As we know, the $SO(10)$ gauge symmetry can be broken down 
to the $SU(3)_C\times SU(2)_L\times U(1)_Y\times U(1)_{\chi}$ gauge symmetry~\cite{gursey, Langacker:2008yv, 
general, PLJW, Erler:2002pr, Kang:2004pp, Kang:2004ix, Kang:2009rd},
where $U(1)_{\chi}$ charges for the SM particles are given in Table~\ref{SO10Qcharge}.

We shall propose the generic $U(1)_X$ model inspired from $SO(10)$,
which is the mixing between the previous $U(1)_{\chi}$ gauge symmetry and $U(1)_Y$ gauge
symmetry. Thus, we have
\begin{eqnarray}
Q_X &=& \cos\alpha \ Q_{Y} + \sin\alpha \ Q_{\chi}~.~\,
\label{SO10-IMIX}
\end{eqnarray}
To break $U(1)_X$ gauge symmetry and give masses to the right-handed
neutrinos, we introduce a SM singlet Higgs $S$ with  $U(1)_{\chi}$ 
charge 10. So the neutrino masses can be explained via the type I seesaw mechanism.
The particles and their quantum numbers under the  
$SU(3)_C\times SU(2)_L\times U(1)_Y\times U(1)_{X}$ gauge symmetry are given in Table \ref{Particle-Spectrum-GSO10}.

\begin{table}[t]
\begin{tabular}{|c|c|c|c|}
\hline
~$Q_i$~ & ~$(\mathbf{3}, \mathbf{2}, \mathbf{1/6}, \cos\alpha/6-\sin\alpha/2{\sqrt{10}}, \mathbf{-1})$~ &
$U_i^c$ &  
~$(\mathbf{\overline{3}}, \mathbf{1}, \mathbf{-2/3},  -2\cos\alpha/3-\sin\alpha/2{\sqrt{10}}, \mathbf{-1})$~ \\
\hline
~$D_i^c$~ & ~$(\mathbf{\overline{3}}, \mathbf{1}, \mathbf{1/3}, \cos\alpha/3+3\sin\alpha/2{\sqrt{10}}, \mathbf{3})$ ~
&~$L_i$~ & ~$(\mathbf{1}, \mathbf{2},  \mathbf{-1/2}, -\cos\alpha/2+3\sin\alpha/2{\sqrt{10}}, \mathbf{3})$~ \\
\hline
$E_i^c$ &  $(\mathbf{1}, \mathbf{1},  \mathbf{1}, \cos\alpha-\sin\alpha/2{\sqrt{10}}, \mathbf{-1})$ &
~$N_i^c$~ &  $(\mathbf{1}, \mathbf{1},  \mathbf{0},-5\sin\alpha/2{\sqrt{10}}, \mathbf{-5})$~ \\
\hline
~$H$~ & ~$(\mathbf{1}, \mathbf{2},  \mathbf{-1/2}, -\cos\alpha/2-2\sin\alpha/2{\sqrt{10}}, \mathbf{-2})$~ &
~$~S$~ &  $(\mathbf{1}, \mathbf{1},  \mathbf{0}, 10\sin\alpha/2{\sqrt{10}},  \mathbf{10})$~     \\
\hline
\end{tabular}
\caption{The particles and their quantum numbers under the
  $SU(3)_C \times SU(2)_L \times U(1)_Y \times U(1)_X$ and $U(1)_{\chi}$ gauge symmetry. Here,
  the correct $U(1)_{\chi}$ charges are the $U(1)_{\chi}$ charges in the above Table divided
  by $2{\sqrt{10}}$.}
\label{Particle-Spectrum-GSO10}
\end{table}

Assuming the interactions between the DM and nucleons are mediated by the $U(1)_X$ gauge boson,
we obtain the coupling ratio $f_n/f_p$  in our model  
\begin{eqnarray}
f_n/f_p &=& \frac{b_u + 2b_d}{2b_u + b_d}  = \frac{(q_Q-q_{U^c})+2(q_Q-q_{D^c})}{2(q_Q-q_{U^c})+(q_Q-q_{D^c})}~,~\,
\label{fnfp}
\end{eqnarray}
where $q_f$ represents the corresponding $U(1)_X$ charge for  particle $f$, 
which is given in Table \ref{Particle-Spectrum-GSO10}. Thus, we have
\begin{eqnarray}
f_n/f_p &=& \frac{\sqrt{10}\cos\alpha-8\sin\alpha}{3\sqrt{10}\cos\alpha-4\sin\alpha}~.~\,
\label{fnfp2}
\end{eqnarray}
To have $f_n/f_p=-0.7$, we obtain $\tan \alpha = \frac{31}{108} \sqrt{10}$. Therefore,
 the $U(1)_X$ charges of the SM particles can be calculated, and are presented 
in the Table \ref{Particle-Spectrum-GSO10-2}. 
To be consistent with GUTs, we choose $g_X\simeq \sqrt{5/3} g_Y$, and assume that
the contribution to the one-loop beta function of $U(1)_X$ from one family of the SM fermions 
in the supersymmetric $U(1)_X$ models
is equal to 2 as in the supersymmetric SMs or GUTs. Thus, we obtain the following 
 normalization factor of $U(1)_X$ charge
\begin{equation}
 N= \sqrt{\frac{Q_{i}^{2}\times 2\times3+U_{i}^{2}\times1\times3+D_{i}^{2}\times1\times3+L_{i}^{2}\times2+E_{i}^{2}\times1+N_{i}^{2}\times1}{2}}~.~\,
\end{equation}
Moreover, we introduce a Dirac fermion $\chi$ as DM candidate, whose  $U(1)_X$ charge is a half integer.
Of course, there are many choices for its $U(1)_X$ charge, and we take 31/2 for simplicity,
which is given in the table as well.

\begin{table}[t]
\begin{tabular}{|c|c|c|c|}
\hline
~$Q_i$~ & ~$(\mathbf{3}, \mathbf{2}, \mathbf{1/6}, \mathbf{1}, \mathbf{-1})$~ &
$U_i^c$ &  
~$(\mathbf{\overline{3}}, \mathbf{1}, \mathbf{-2/3}, \mathbf{-35},  \mathbf{-1})$~ \\
\hline
~$D_i^c$~ & ~$(\mathbf{\overline{3}}, \mathbf{1}, \mathbf{1/3}, \mathbf{33}, \mathbf{3})$ ~
&~$L_i$~ & ~$(\mathbf{1}, \mathbf{2},  \mathbf{-1/2}, \mathbf{-3}, \mathbf{3})$~ \\
\hline
$E_i^c$ &  $(\mathbf{1}, \mathbf{1},  \mathbf{1}, \mathbf{37}, \mathbf{-1})$ &
~$N_i^c$~ &  $(\mathbf{1}, \mathbf{1},  \mathbf{0}, \mathbf{-31},\mathbf{-5})$~ \\
\hline
~$H$~ & ~$(\mathbf{1}, \mathbf{2},  \mathbf{-1/2}, \mathbf{-34}, \mathbf{-2})$~ &
~$~S$~ &  $(\mathbf{1}, \mathbf{1},  \mathbf{0},  \mathbf{62}, \mathbf{10})$~     \\
\hline
~$~\chi$~ & ~$(\mathbf{1}, \mathbf{1},  \mathbf{0}, \mathbf{31/2}, \mathbf{5/2})$~ &
~$ $~ &  $ $~     \\
\hline
\end{tabular}
\caption{The particles and their quantum numbers under the
  $SU(3)_C \times SU(2)_L \times U(1)_Y \times U(1)_X$ and $U(1)_{\chi} $ gauge symmetry. Here,
  the correct $U(1)_{\chi}$ and $U(1)_X$ charges are their charges in the above table divided
  by $2{\sqrt{10}}$ and $2 \sqrt{1162}$.}
\label{Particle-Spectrum-GSO10-2}
\end{table}

The Lagarangian is given by
\begin{eqnarray}
  -{\cal L} &=& m_S^2 |S|^2 + m_H^2 |H|^2 + \frac{\lambda_S}{2} |S|^4 +  \frac{\lambda_H}{2} |H|^4
+  {\lambda_{SH}} |S|^2 |H|^2 + \left( y_{ij}^U Q_i U_j^c {\tilde H} 
\right. \nonumber\\&& \left.
+ y_{ij}^D Q_i D_j^c H + y_{ij}^E L_i E_j^c H
  + y_{ij}^{N} L_i N_j^c {\tilde H} + y_{ij}^{MN} S N_i^c N_j^c + {\rm H. C.}\right)~,~
  \label{lag:yukawa-GSO10}
\end{eqnarray}
where ${\tilde H}= \mathrm{i}\sigma_2 H^* $.

We parametrize the Higgs fields as follows:
\begin{align}
S=v_s+S_1+\mathrm{i}S_2~ ~,   H &= \begin{pmatrix} &H^+& \\&v_h+H_1+ \mathrm{i}H_2& \end{pmatrix}~.
\end{align}
After gauge symmetry breakings, the vacuum expectation values (VEVs) of these Higgs fields are given by
\begin{align}
<S>= v_s~ ~ , <H> &= \begin{pmatrix} 0 \\ v_h \end{pmatrix}.
\end{align}

We have four Nambu-Goldstone bosons from $H^{\pm}$, $H_2$, and $S_2$, as well as
two neutral physical scalars  $s$ and $h$  from the mixings of $S_1$ and $H_1$ 
via their following mass matrix 
\begin{align}
M_{(S_1,~H_1)} =
\begin{bmatrix}
2v_s^2\lambda_S\   & 2v_hv_s\lambda_{SH} \\
2v_hv_s\lambda_{SH} &2v_h^2\lambda_H\\
\end{bmatrix}~.~\,
\end{align}

The physical scalars $s$ and $h$ can be written as the linear combination of $H_1$ and $S_1$
\begin{align}
\begin{pmatrix}  &s& \\ &h& \end{pmatrix}&=
\begin{pmatrix}
\cos \theta   & -\sin \theta \\
\sin \theta &\cos \theta\\
\end{pmatrix}
\begin{pmatrix}  &S_1& \\ &H_1& \end{pmatrix}~,~\,
\end{align}
where the mixing angle $\theta$ is
\begin{align}
\tan 2\theta=\frac{2v_hv_s\lambda_{SH}}{v_h^2\lambda_H-v_s^2\lambda_S}~.~\,
\end{align}
The Higgs masses are, respectively,
\begin{align}
m_{s/h}^2=\lambda _h v_h^2+\lambda _s v_s^2 \pm \sqrt{4 \lambda _{sh}^2 v_h^2 v_s^2-2 \lambda _s \lambda _h v_h^2 v_s^2+\lambda _h^2 v_h^4+\lambda _s^2 v_s^4}.
\end{align}
Because $h$ is the SM Higgs field, we should have $v_h=174$ GeV and $m_h=125$ GeV.

Next we shall discuss the gauge boson masses. The covariant derivative in our model is
\begin{align}
D_\mu &=\partial_\mu-\mathrm{i}g_2T^iA_\mu^i-\mathrm{i}g_Y Y B_\mu-\mathrm{i}g_X X C_\mu \\
&=\partial_\mu-\mathrm{i}\left(
\begin{array}{cc}
 \frac{1}{2} g_2A_\mu^3-\frac{1}{2}g_YB_\mu+g_XXC_\mu &  \frac{g_2 W_\mu^+}{\sqrt{2}} \\
 \frac{g_2 W_\mu^-}{\sqrt{2}} & -\frac{1}{2}g_2A_\mu^3-\frac{1}{2}g_YB_\mu+g_XXC_\mu \\
\end{array}
\right) ~,~\,
\nonumber
\end{align}
where $i=1,2,3$, $T^i$ are the three generators of $SU(2)_L$, 
$Y$ and $X$ are the charges, respectively for $U(1)_Y$ and $U(1)_X$, 
$g_2$, $g_Y$ and $g_X$ are the gauge couplings, respectively, for $SU(2)_L$, $U(1)_Y$, and $U(1)_X$, 
$W^\pm _\mu$ are the SM charged gauge bosons $W^\pm _\mu=\frac{A^1_\mu \mp \mathrm{i}A^2_\mu}{\sqrt{2}}$,
$A^i_\mu$ and $B_\mu$ are the SM gauge fields, and $C_\mu$ represents for the $U(1)_X$ gauge field.

After gauge symmetry breaking, we obtain the gauge boson masses from the kinetic terms of the Higgs fields 
\begin{align}
{\cal L}_{GM}&=(D_\mu H)^\dagger D_\mu H +(D_\mu S)^\dagger D_\mu S \\
&\overset{\rm VEV}{=} v_h^2 (-\frac{1}{2}g_2A_\mu^3 -\frac{1}{2}g_YB_\mu +g_XX_h C_\mu)^2 +
v_s^2 g_X^2 C_\mu^2 X_s^2+ \frac{1}{2}g_2^2 v_h^2 W_\mu^+W^{-\mu} ~.~\nonumber
\end{align}

The gauge boson mass matrix in the basis ($A_\mu^3$ , $B_\mu$, $C_\mu$) is 
\begin{align}
\left(
\begin{array}{ccc}
 \frac{1}{2}g_2^2 v_h^2 & \frac{1}{2}g_2 g_Y v_h^2 & -g_2g_Xv_h^2 X_h \\
 \frac{1}{2}g_2 g_Y v_h^2 & \frac{1}{2}g_Y^2 v_h^2 & -g_Y g_X v_h^2 X_h \\
 -g_2 g_X v_h^2 X_h & -g_Y g_X v_h^2 X_h & 2g_X^2 v_h^2 X_h^2+ 2g_X^2 v_s^2 X_s^2 \\
\end{array}
\right)~,~\,
\end{align}
and then the gauge boson mass are given by 
\begin{align}
m_1^2&=0 ~ ~, m_{2,3}^2=\frac{1}{4} \left(g_2^2 v_h^2 + g_Y^2 v_h^2 + 4g_X^2 v_h^2 X_h^2 + 
   4g_X^2 v_s^2 X_s^2 \right.\nonumber \\
   &\left.\pm \sqrt{(-g_2^2 v_h^2 - g_Y^2 v_h^2 - 4g_X^2 v_h^2 X_h^2 - 
        4g_X^2 v_s^2 X_s^2)^2 - 
      4 (4g_2^2 g_X^2 v_h^2 v_s^2 X_s^2 + 4g_Y^2 g_X^2 v_h^2 v_s^2 X_s^2)}\right)~,~
\end{align}
which obviously are the masses for photon, $Z$, and $Z'$ gauge bosons, respectively. 
Because $Z'$ is much heavier than $Z$,  we have $m_3^2=m_{Z}^2$.
With these calculations, we can give these parameters the proper values to 
satisfy the experimental constraints. For example, we can have
 $v_s=8000$ GeV, $m_{H}=1884$ GeV,$m_{S}=3465$ GeV,$\lambda_S=0.17$, $\lambda_H=0.27$, $\lambda_{SH}=0.05$, $g_{X}=0.50$, 
and then get $m_h=125$ GeV,$m_s=4900$ GeV, and $m_{Z'}=5500$ GeV. 

At low energy, we have the SM particles, $Z'$, and DM $\chi$. Therefore,
 we can use the simplified $U(1)_X$ model whose the interactions are given by
\begin{eqnarray}
-{\cal L} = \sum_q g_{u} \bar{u} \gamma^{\mu}uZ_{\mu}^\prime + g_{uA} \bar{u} \gamma^{\mu}\gamma^{5}uZ_{\mu}^\prime + g_{d} \bar{d} \gamma^{\mu}dZ_{\mu}^\prime + g_{dA} \bar{d} \gamma^{\mu}\gamma^{5}dZ_{\mu}^\prime  +g_{\chi} \bar{\chi} \gamma^{\mu}\chi Z_{\mu}^\prime ~,~
\end{eqnarray}
Where $g_f=g_Xq_f$ with $q_f$  the $U(1)_X$ charge for the fermion $f$, 
and we have
\begin{eqnarray}
\label{eq:rate}
g_u:g_d:g_{uA}:g_{dA}:g_{\chi}=36:-32:34:-34:31,~q_u=18/\sqrt{1162}~.~\,
\end{eqnarray}

\section{The $U(1)_X$ Model with the IVDM $\chi$}\label{sec:exp}

First, we would like to study the LHC constraints on the $Z'$ mass via the code $Z'$ explorer~\cite{Alvarez:2020yim}.
For simplicity, we take $g_X=\sqrt{\frac{5}{3}}g_Y=0.46$, and then we have $g_{uL}=g_{dL} \approx 0.0067$, 
$g_{uR} \approx 0.236$, and $g_{dR} \approx -0.223$. Further parameter settings can be found in the Appendix~\ref{app:width}. 
Because the $U(1)_X$ charge of $Q_i$ is much smaller than $U_i^c$ and $D_i^c$, the coupling 
between the left-handed quarks and $Z'$ is much smaller than the right-handed quarks. 

\begin{figure}[ht]
	\centering
	\includegraphics[width=0.5\textwidth]{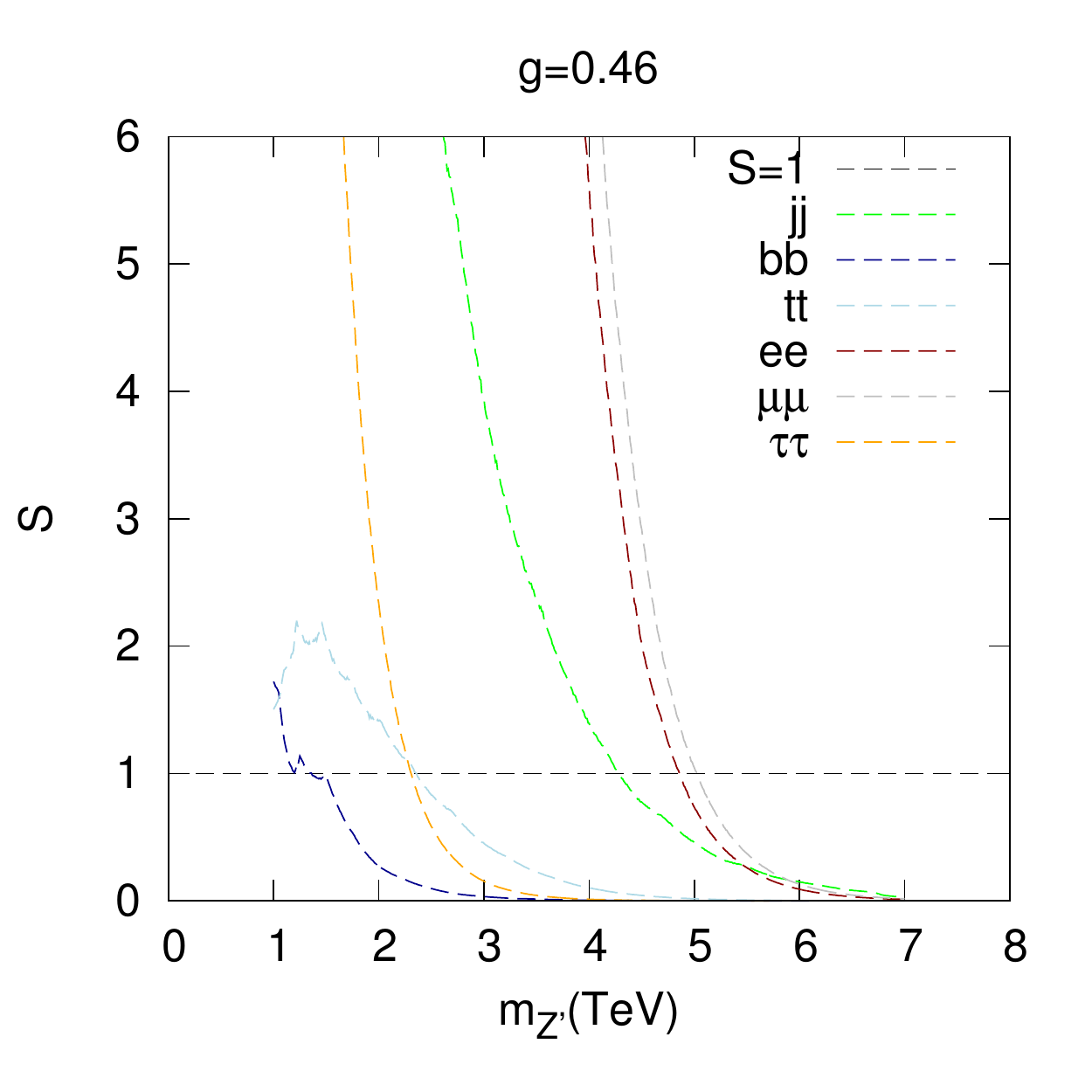}
	\hspace{3em} 
	\caption{The signal strengths of the SM fermion final states versus $Z'$ mass
for the $Z'$ searches at the LHC. The LHC bound on the $Z'$ mass is about 5.03 TeV.} 
	\label{LHC}
\end{figure}

We present the LHC simulation results in Fig.~\ref{LHC}. $S=\frac{\sigma_{pred}}{\sigma_{lim}}$ is the signal strength  
for each channel, where $\sigma_{pred}$ is the predicted  $Z^{\prime}$ production cross section times branching ratio times acceptance and $\sigma_{lim}$ is the corresponding predicted experimental upper limit at the $95\%$ confidence level (C.L).  If $S > 1$, then the corresponding point in the parameter space is experimentally excluded. 
If $S < 1$ for all channels, then 
the corresponding point is viable. Therefore, the  channels with the $e\bar{e}$ and $\mu\bar{\mu}$ final states give
the strong constraints on $Z'$ mass, and we obtain that the low bound on  $Z'$ mass in our model is  
around $5.03$ TeV. Because  the current LHC mass bound on  generic $Z'$ is about $5$ TeV~\cite{Aad:2019fac},
our numerical result is consistent with LHC searches.
For simplicity,  we shall take $M_{Z'}=5.5$ TeV in the following study.

\begin{figure}[ht]
	\centering
	\begin{subfigure}{0.45\textwidth} 
		\includegraphics[width=\textwidth]{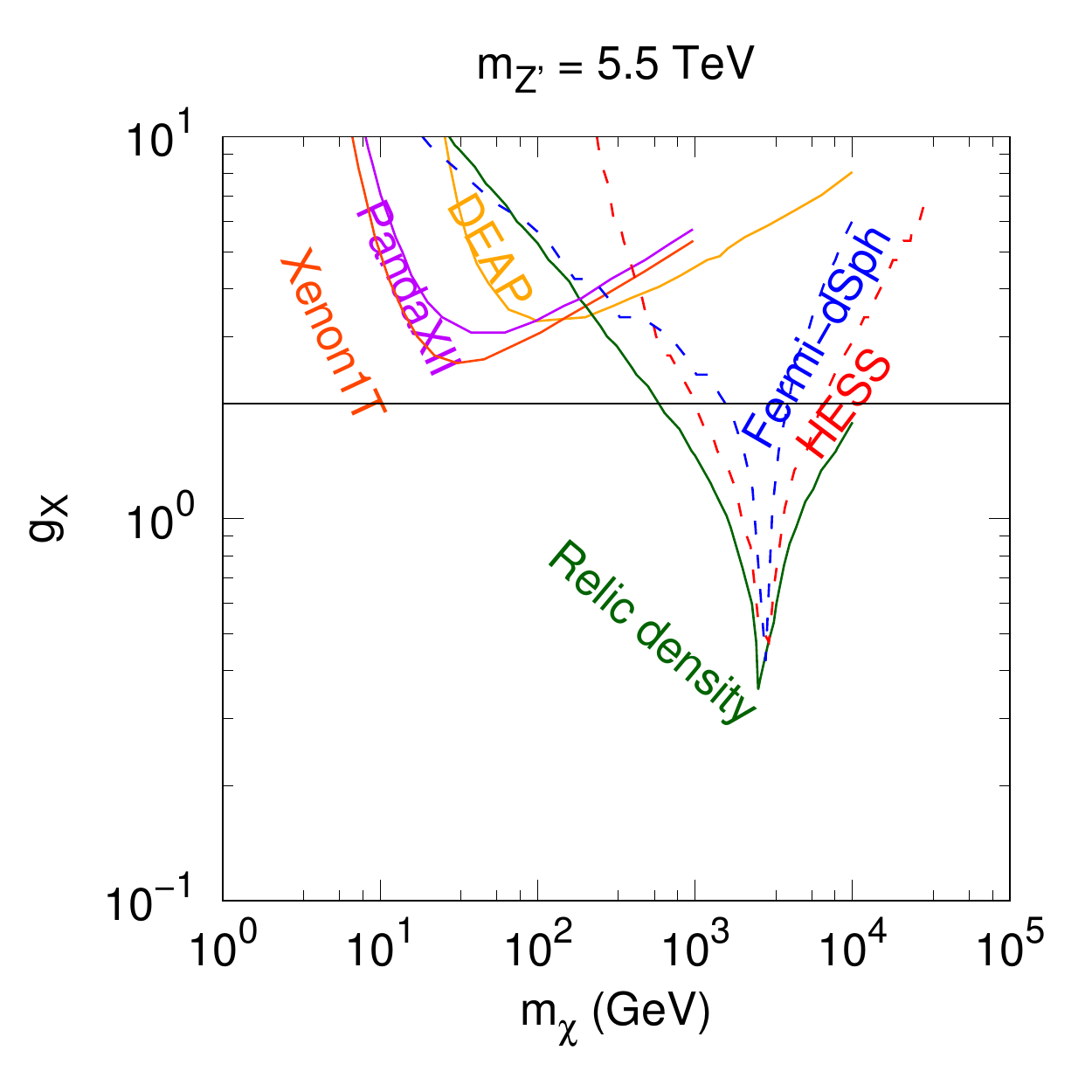}
		\caption{$m_{Z'}=5.5$ TeV} 
	\end{subfigure}
	\hspace{3em} 
	\begin{subfigure}{0.45\textwidth} 
		\includegraphics[width=\textwidth]{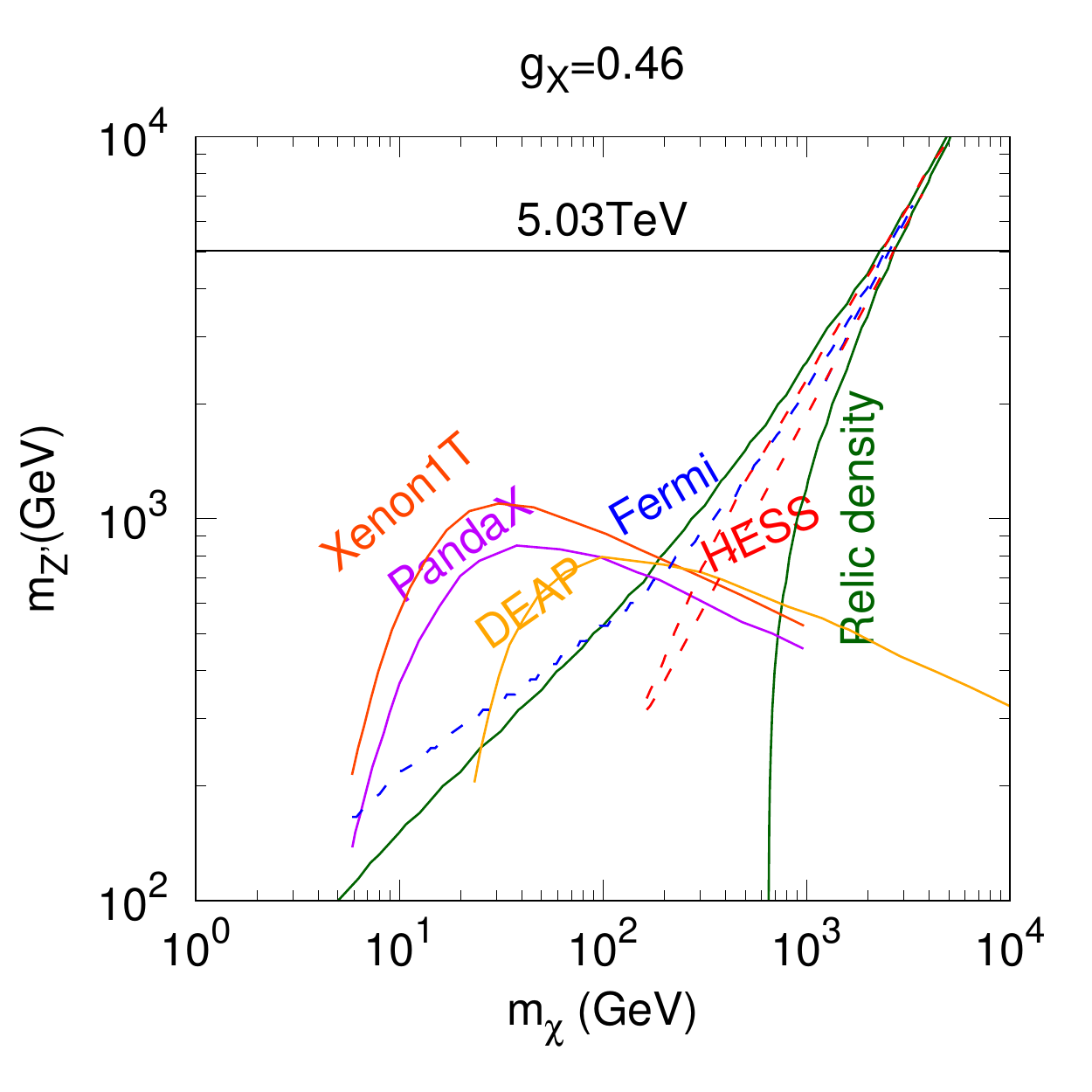}
		\caption{$g_X=0.46$} 
	\end{subfigure}
	\caption{Exclusion line for direct detection experiments and indirect detection
experiments.} 
	\label{chi}
\end{figure}

Second, we shall consider the direct and indirect experimental constraints on the DM $\chi$,
and present the simulation results
 in the $g_X$ versus $m_{\chi}$ plane
with $M_{Z'}=5.5$ TeV and in the $M_{Z'}$ versus $m_{\chi}$ plane with $g_X=0.46$, 
respectively, in the left and right panels of Fig.~\ref{chi}.
In the left panel, to have the decay width of $Z'$ smaller than $Z'$ mass from
the $Z'$ particle point of view,  we obtain that the upper bound on $g_X$ is roughly 2, 
{\it i.e.}, $g_X < 2$. In the right panel, we have the low bound on $Z'$ mass 
from the LHC constraints, {\it i.e.}, $m_{Z'} > 5.03$ TeV.
 These two conditions are shown as the solid black lines in Fig.~\ref{chi}.
Also, the dark-green line indicates the parameter space with the observed 
dark matter relic density, {\it i.e.}, $\Omega_{\chi}h^2=0.12$ and the relic density is calculated by the popular code MicrOMEGAs~\cite{Belanger:2006is,Belanger:2013oya}.

The solid purple, orange and yellow lines, respectively, correspond to the constraints from 
the PandaX-II (2017)~\cite{Cui:2017nnn}, 
Xenon1T (2018)~\cite{Aprile:2018dbl}, and DEAP3600 (2019)~\cite{Ajaj:2019imk} experiments. 
The first two experiments are xenon-based DM direct detections, while the last one is argon-based. 
On account of isospin-violating affects, they have the rescale factors 
 around 7600 and 235~\cite{Yaguna:2016bga}, respectively. Compared to the two black lines, 
the DM direct detection experiments barely give additional constraints, 
and then we have escaped these experimental constraints.

The dashed blue and red lines correspond to the constraints from the Fermi-dSph (6 year)~\cite{TheFermi-LAT:2017vmf} 
and HESS (254h)~\cite{Abdallah:2016ygi}, respectively.  In addition, we should clarify that the DAMA and GoGeNT are not shown, since they have less constrains.
The interesting parameter spaces for our simulations are distributed in the resonant regions,
which have $m_\chi \sim \frac{1}{2}m_{Z'}$, and these regions together with the regions below the black lines
constitute the main parameter spaces in our model.

Comparing  the direct and indirect experiments, we find that the 
indirect detection have much better sensitivity near the  resonant regions 
with $m_\chi \sim \frac{1}{2}m_{Z'} $ due to resonant enhancement. Beyond these regions, 
the direct detection experiments have better sensitivities. 
In short, there are still some viable parameter spaces in our $U(1)_X$ model.

\section{The Generic $U(1)_X$ Models with LHC Bounds on the $Z'$ Masses}\label{Neutrino}

In the following,
we shall consider four scenarios and study the LHC low bounds on the $Z'$ mass.

\subsection{Scenario I: The $U(1)_X$ model with zero $U(1)_X$ charge for the quark doublets $Q_i$ }

\begin{figure}[htb]
	\centering
	\includegraphics[width=0.5\textwidth]{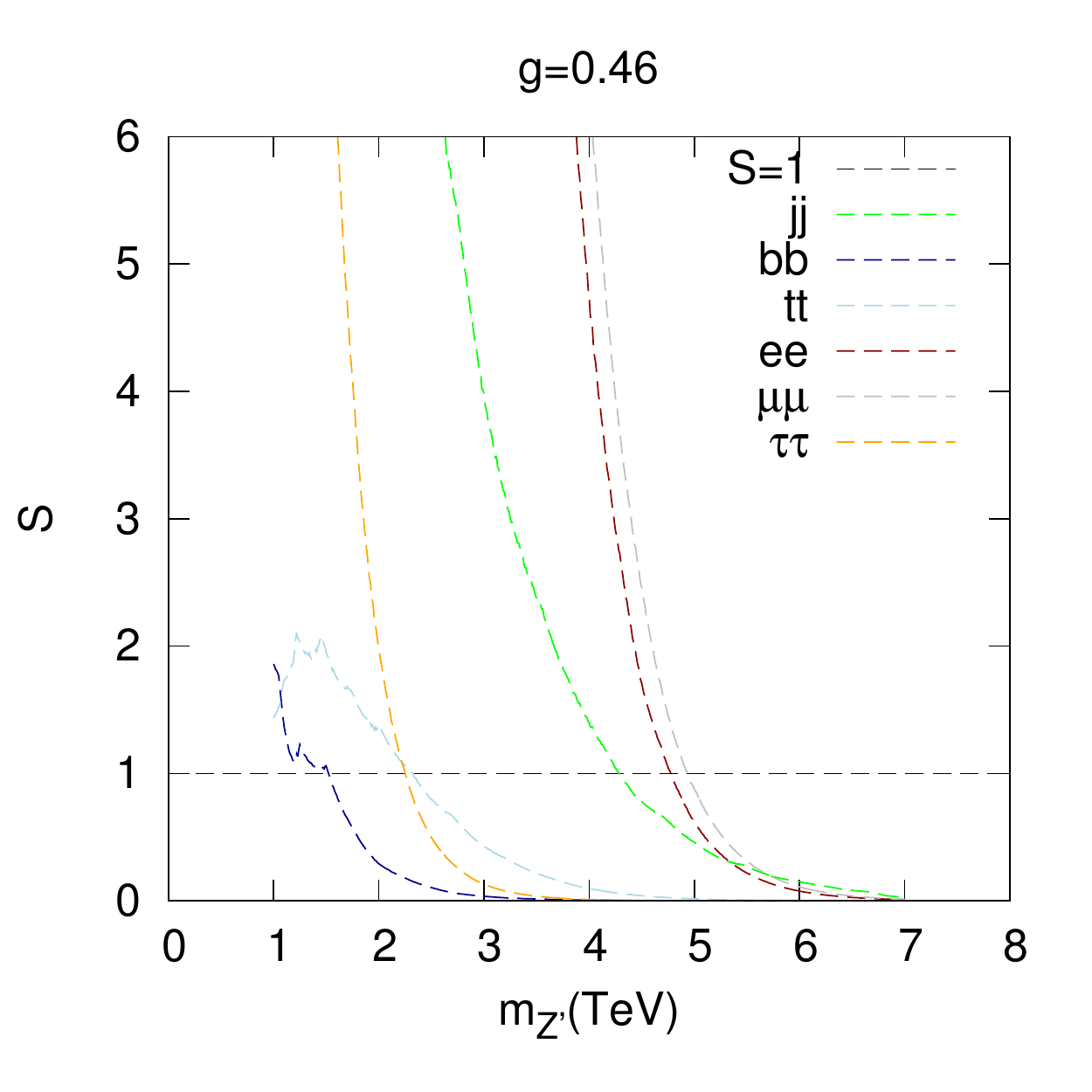}
	\hspace{3em} 
	\caption{The signal strengths of the SM fermion final states versus $Z'$ mass
for the $Z'$ searches at the LHC in scenario I. The low bound on $Z'$ mass is around 4.94 TeV.} 
	\label{Scenario1}
\end{figure}

In scenario I, we consider that the $U(1)_X$ charge for the quark doublets $Q_i$ is equal to 0,
and then we obtain 
\begin{eqnarray}
\tan \alpha = \frac{{\sqrt{10}}}{3} ~.~
\end{eqnarray}
The particles and their quantum numbers under the  
$SU(3)_C\times SU(2)_L\times U(1)_Y\times U(1)_{X}$ gauge symmetry are given 
in Table \ref{Particle-Spectrum-GSO10-3}. 
We present the LHC simulation results in Fig.~\ref{Scenario1},
and obtain that low bound on $Z'$ boson mass is about 4.94 TeV.

\begin{table}[ht]
  \begin{tabular}{|c|c|c|c|}
  \hline
  ~$Q_i$~ & ~$(\mathbf{3}, \mathbf{2}, \mathbf{1/6}, \mathbf{0}, \mathbf{-1})$~ &
  $U_i^c$ &  
  ~$(\mathbf{\overline{3}}, \mathbf{1}, \mathbf{-2/3}, \mathbf{1},  \mathbf{-1})$~ \\
  \hline
  ~$D_i^c$~ & ~$(\mathbf{\overline{3}}, \mathbf{1}, \mathbf{1/3}, \mathbf{-1}, \mathbf{3})$ ~
  &~$L_i$~ & ~$(\mathbf{1}, \mathbf{2},  \mathbf{-1/2}, \mathbf{0}, \mathbf{3})$~ \\
  \hline
  $E_i^c$ &  $(\mathbf{1}, \mathbf{1},  \mathbf{1}, \mathbf{-1}, \mathbf{-1})$ &
  ~$N_i^c$~ &  $(\mathbf{1}, \mathbf{1},  \mathbf{0}, \mathbf{1},\mathbf{-5})$~ \\
  \hline
  ~$H$~ & ~$(\mathbf{1}, \mathbf{2},  \mathbf{-1/2}, \mathbf{1}, \mathbf{-2})$~ &
  ~$~S$~ &  $(\mathbf{1}, \mathbf{1},  \mathbf{0},  \mathbf{-2}, \mathbf{10})$~     \\
  \hline
\end{tabular}
\caption{The particles and their quantum numbers under the
  $SU(3)_C \times SU(2)_L \times U(1)_Y \times U(1)_X$ and $U(1)_{\chi}$ gauge symmetry in scenario I. Here,
  the correct $U(1)_{\chi}$ and $U(1)_X$ charges are their charges in the above table divided
  by $2{\sqrt{10}}$ and $2$.}
\label{Particle-Spectrum-GSO10-3}
\end{table}

\subsection{Scenario II: The $U(1)_X$ model with zero $U(1)_X$ charge for the right-handed up-type quarks $U^c_i$ }

\begin{figure}[htb]
	\centering
		\includegraphics[width=0.5\textwidth]{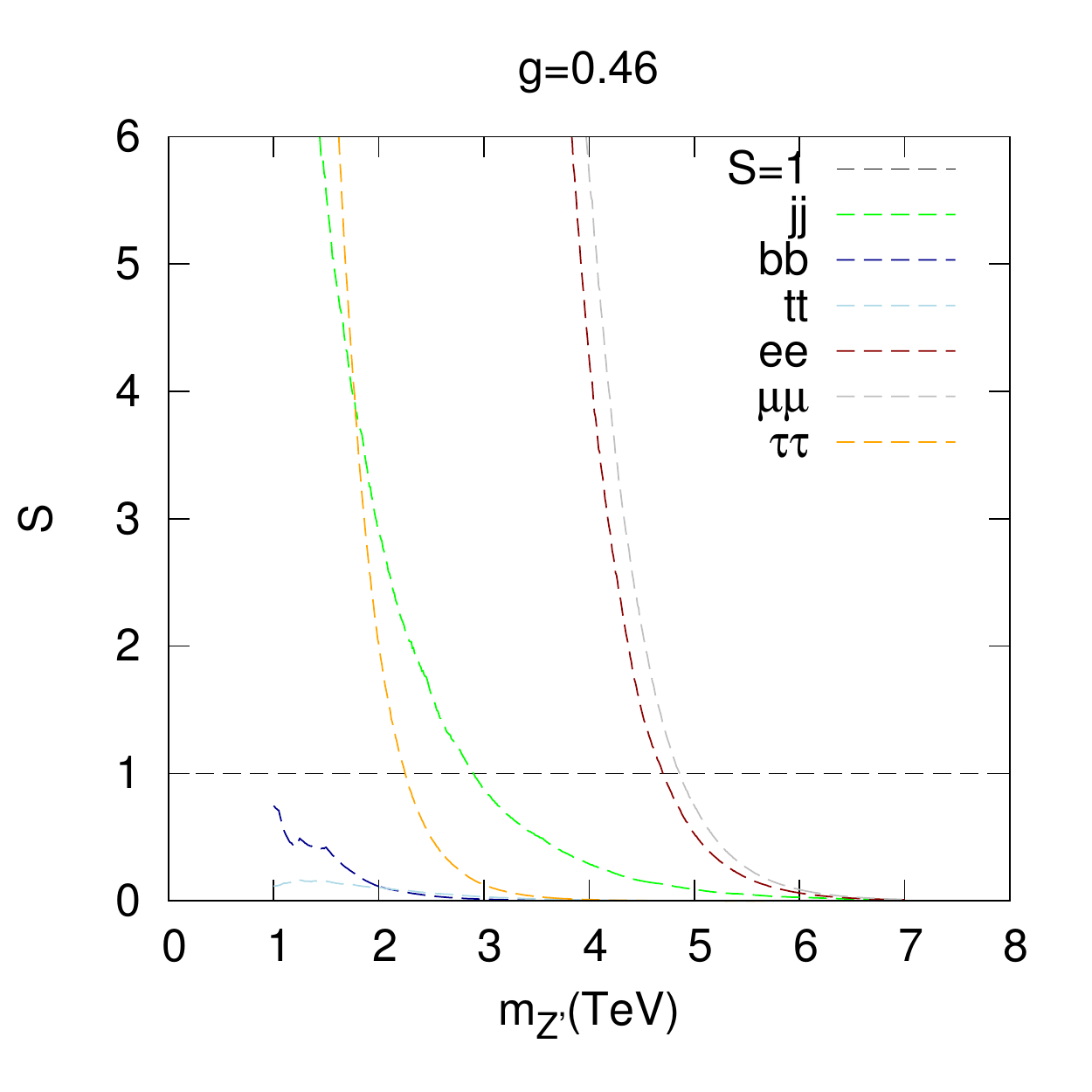}
	\hspace{3em} 
	\caption{The signal strengths of the SM fermion final states versus $Z'$ mass
for the $Z'$ searches at the LHC in scenario II. The low bound on $Z'$ mass is around 4.87 TeV.} 
	\label{Scenario2}
\end{figure}

In scenario II, we consider that the $U(1)_X$ charge for the right-handed up-type quarks $U^c_i$ is equal to 0,
and then we obtain 
\begin{eqnarray}
\tan \alpha = - \frac{4 {\sqrt{10}}}{3} ~.~
\end{eqnarray}
The particles and their quantum numbers under the  
$SU(3)_C\times SU(2)_L\times U(1)_Y\times U(1)_{X}$ gauge symmetry are given 
in Table \ref{Particle-Spectrum-GSO10-4}. 
We present the LHC simulation results in Fig.~\ref{Scenario2},
and obtain that low bound on $Z'$ boson mass is  about 4.87  TeV.

\begin{table}[ht]
  \begin{tabular}{|c|c|c|c|}
  \hline
  ~$Q_i$~ & ~$(\mathbf{3}, \mathbf{2}, \mathbf{1/6}, \mathbf{1}, \mathbf{-1})$~ &
  $U_i^c$ &  
  ~$(\mathbf{\overline{3}}, \mathbf{1}, \mathbf{-2/3}, \mathbf{0},  \mathbf{-1})$~ \\
  \hline
  ~$D_i^c$~ & ~$(\mathbf{\overline{3}}, \mathbf{1}, \mathbf{1/3}, \mathbf{-2}, \mathbf{3})$ ~
  &~$L_i$~ & ~$(\mathbf{1}, \mathbf{2},  \mathbf{-1/2}, \mathbf{-3}, \mathbf{3})$~ \\
  \hline
  $E_i^c$ &  $(\mathbf{1}, \mathbf{1},  \mathbf{1}, \mathbf{2}, \mathbf{-1})$ &
  ~$N_i^c$~ &  $(\mathbf{1}, \mathbf{1},  \mathbf{0}, \mathbf{4},\mathbf{-5})$~ \\
  \hline
  ~$H$~ & ~$(\mathbf{1}, \mathbf{2},  \mathbf{-1/2}, \mathbf{1}, \mathbf{-2})$~ &
  ~$~S$~ &  $(\mathbf{1}, \mathbf{1},  \mathbf{0},  \mathbf{-8}, \mathbf{10})$~     \\
  \hline
\end{tabular}
\caption{The particles and their quantum numbers under the
  $SU(3)_C \times SU(2)_L \times U(1)_Y  \times U(1)_X$ and $U(1)_{\chi}$ gauge symmetry in scenario II. Here,
  the correct $U(1)_{\chi}$ and $U(1)_X$ charges are their charges in the above table divided
  by $2{\sqrt{10}}$ and $2\sqrt{7}$.}
\label{Particle-Spectrum-GSO10-4}
\end{table}

\subsection{Scenario III: The $U(1)_X$ model with zero $U(1)_X$ charge for the right-handed down-type quarks $D^c_i$ }

\begin{figure}[htb]
	\centering
	\includegraphics[width=0.5\textwidth]{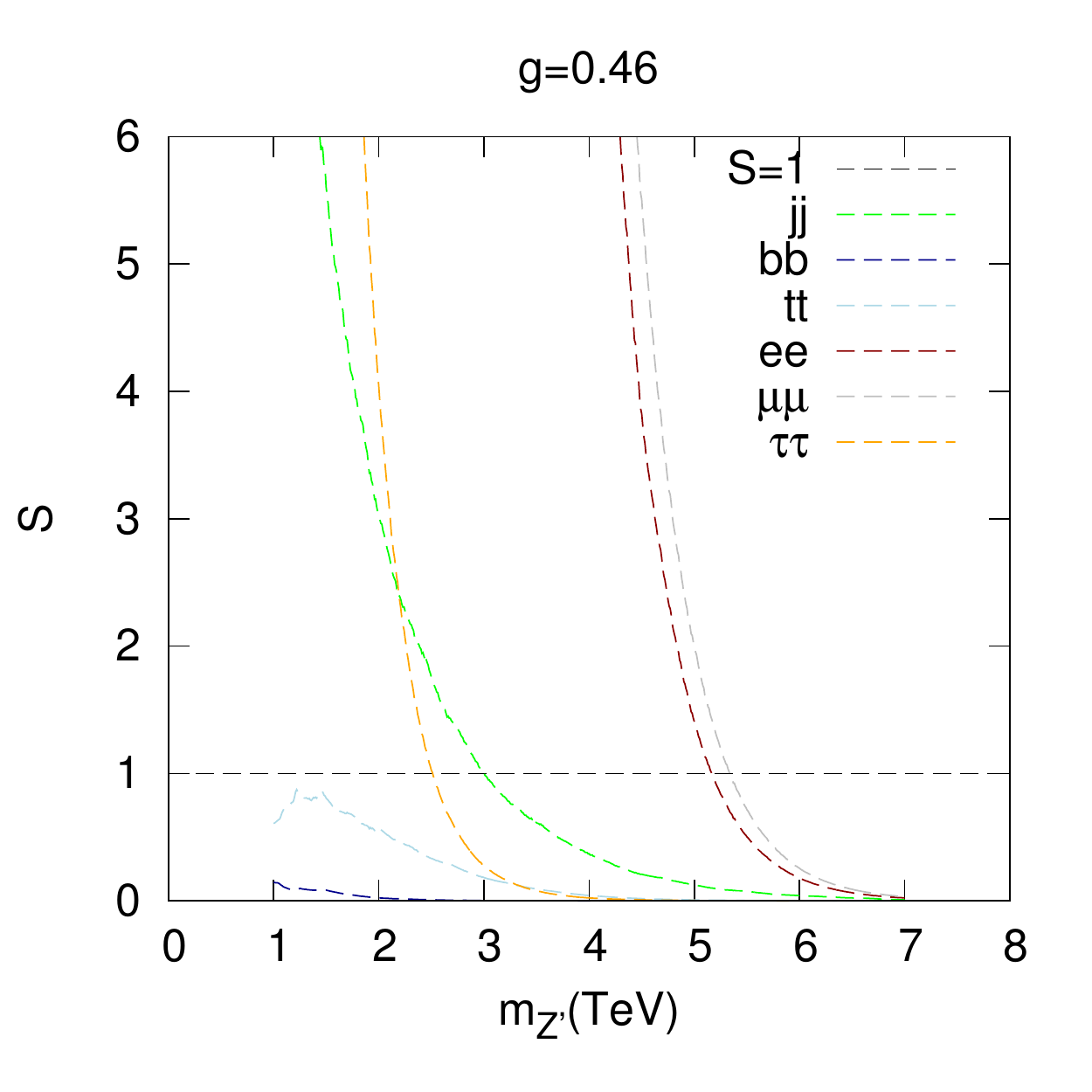}
	\hspace{3em} 
	\caption{The signal strengths of the SM fermion final states versus $Z'$ mass
for the $Z'$ searches at the LHC in scenario III. The low bound on $Z'$ boson mass is around 5.34 TeV.} 
	\label{Scenario3}
\end{figure}

In scenario III, we consider that the $U(1)_X$ charge for the right-handed down-type quarks $D^c_i$ is equal to 0,
and then we obtain 
\begin{eqnarray}
\tan \alpha = - \frac{2 {\sqrt{10}}}{9} ~.~
\end{eqnarray}
The particles and their quantum numbers under the  
$SU(3)_C\times SU(2)_L\times U(1)_Y\times U(1)_{X}$ gauge symmetry are given 
in Table \ref{Particle-Spectrum-GSO10-5}. 
We present the LHC simulation results in Fig.~\ref{Scenario3},
and obtain that the low bound on the $Z'$ boson mass is about 5.34  TeV.

\begin{table}[ht]
  \begin{tabular}{|c|c|c|c|}
  \hline
  ~$Q_i$~ & ~$(\mathbf{3}, \mathbf{2}, \mathbf{1/6}, \mathbf{1}, \mathbf{-1})$~ &
  $U_i^c$ &  
  ~$(\mathbf{\overline{3}}, \mathbf{1}, \mathbf{-2/3}, \mathbf{-2},  \mathbf{-1})$~ \\
  \hline
  ~$D_i^c$~ & ~$(\mathbf{\overline{3}}, \mathbf{1}, \mathbf{1/3}, \mathbf{0}, \mathbf{3})$ ~
  &~$L_i$~ & ~$(\mathbf{1}, \mathbf{2},  \mathbf{-1/2}, \mathbf{-3}, \mathbf{3})$~ \\
  \hline
  $E_i^c$ &  $(\mathbf{1}, \mathbf{1},  \mathbf{1}, \mathbf{4}, \mathbf{-1})$ &
  ~$N_i^c$~ &  $(\mathbf{1}, \mathbf{1},  \mathbf{0}, \mathbf{2},\mathbf{-5})$~ \\
  \hline
  ~$H$~ & ~$(\mathbf{1}, \mathbf{2},  \mathbf{-1/2}, \mathbf{-1}, \mathbf{-2})$~ &
  ~$~S$~ &  $(\mathbf{1}, \mathbf{1},  \mathbf{0},  \mathbf{-4}, \mathbf{10})$~     \\
  \hline
\end{tabular}
\caption{The particles and their quantum numbers under the
  $SU(3)_C \times SU(2)_L \times U(1)_Y  \times U(1)_X$ and $U(1)_{\chi}$ gauge symmetry in scenario III. Here,
  the correct $U(1)_{\chi}$ and $U(1)_X$ charges are their charges in the above table divided
  by $2{\sqrt{10}}$ and $2\sqrt{7}$.}
\label{Particle-Spectrum-GSO10-5}
\end{table}

\subsection{Scenario IV: The $U(1)_X$ model with approximately equal
 $U(1)_X$ charges for the quark doublets $Q_i$, right-handed up-type quarks $U^c_i$, and right-handed down-type quarks $D^c_i$ }

\begin{figure}[htb]
	\centering
	\includegraphics[width=0.5\textwidth]{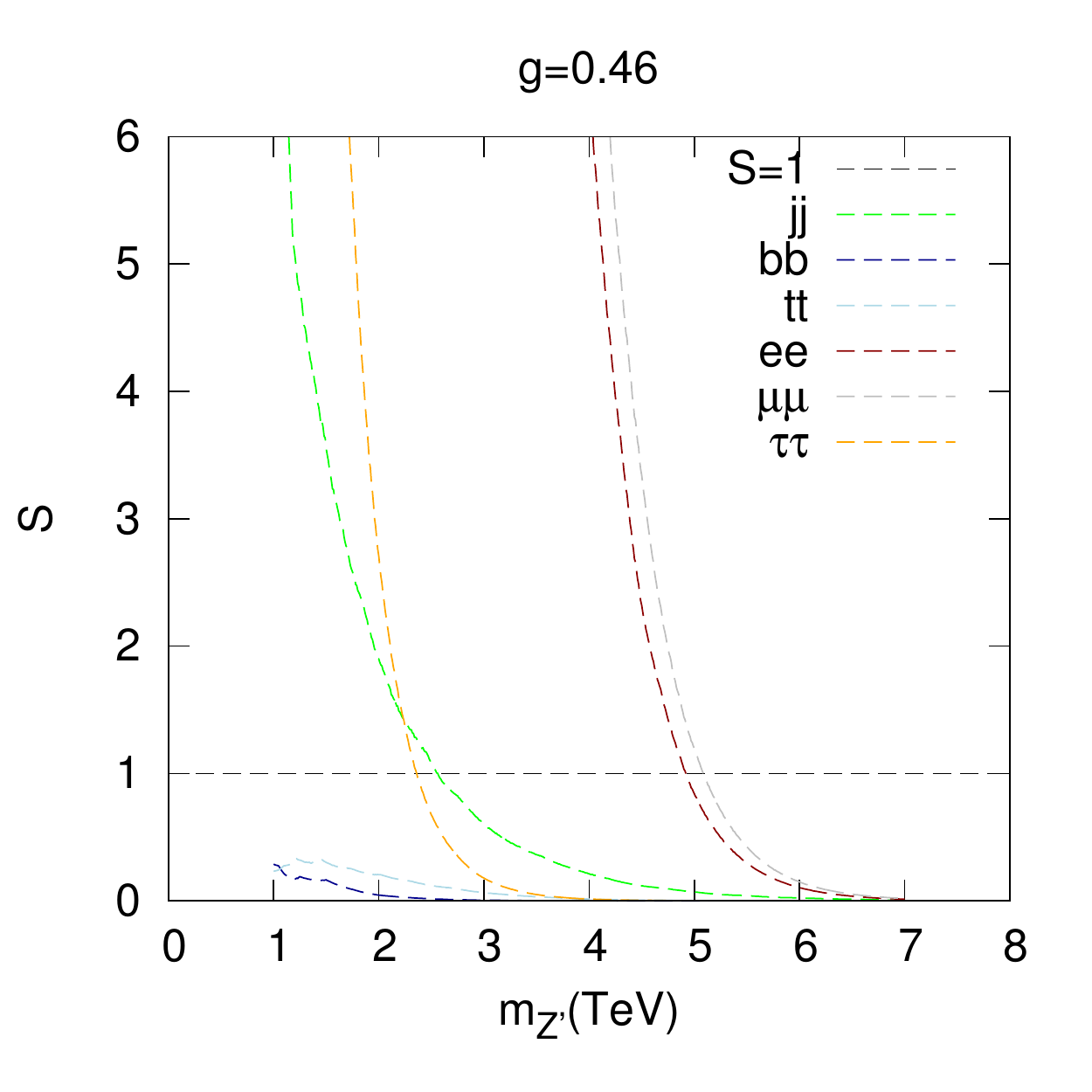}
	\hspace{3em} 
	\caption{The signal strengths of the SM fermion final states versus $Z'$ mass
for the $Z'$ searches at the LHC in scenario IV. The low bound on $Z'$ mass 
is around 5.09 TeV.} 
	\label{Scenario4}
\end{figure}

In scenario IV, we consider that the $U(1)_X$ charge for the quark doublets $Q_i$, 
right-handed up-type quarks $U^c_i$ and right-handed down-type quarks $D^c_i$ 
are approximately equal, and then we obtain 
\begin{eqnarray}
  \tan \alpha = - \sqrt{\frac{13}{5}} ~.~
\end{eqnarray}
The particles and their quantum numbers under the  
$SU(3)_C\times SU(2)_L\times U(1)_Y\times U(1)_{X}$ gauge symmetry are given 
in Table \ref{Particle-Spectrum-GSO10-6}. 
We present the LHC simulation results in Fig.~\ref{Scenario4},
and obtain that low bound on $Z'$ boson mass is about 5.09  TeV.
Therefore, scenario II can relax the LHC $Z'$ mass bound a little bit.

\begin{table}[ht]
  \begin{tabular}{|c|c|c|c|}
  \hline
  ~$Q_i$~ & ~$(\mathbf{3}, \mathbf{2}, \mathbf{1/6}, \mathbf{1}, \mathbf{-1})$~ &
  $U_i^c$ &  
  ~$(\mathbf{\overline{3}}, \mathbf{1}, \mathbf{-2/3},  \mathbf{\frac{317-75\sqrt{26}}{67}}, \mathbf{-1})$~ \\
  \hline
  ~$D_i^c$~ & ~$(\mathbf{\overline{3}}, \mathbf{1}, \mathbf{1/3}, \mathbf{\frac{75\sqrt{26}-451}{67}}, \mathbf{3})$ ~
  &~$L_i$~ & ~$(\mathbf{1}, \mathbf{2},  \mathbf{-1/2}, \mathbf{-3}, \mathbf{3})$~ \\
  \hline
  $E_i^c$ &  $(\mathbf{1}, \mathbf{1},  \mathbf{1}, \mathbf{\frac{75\sqrt{26}-183}{67}}, \mathbf{-1})$ &
  ~$N_i^c$~ &  $(\mathbf{1}, \mathbf{1},  \mathbf{0},\mathbf{\frac{585-75\sqrt{26}}{67}}, \mathbf{-5})$~ \\
  \hline
  ~$H$~ & ~$(\mathbf{1}, \mathbf{2},  \mathbf{-1/2}, \mathbf{\frac{384-75\sqrt{26}}{67}}, \mathbf{-2})$~ &
  ~$~S$~ &  $(\mathbf{1}, \mathbf{1},  \mathbf{0}, \mathbf{\frac{150\sqrt{26}-1170}{67}},  \mathbf{10})$~     \\
  \hline
\end{tabular}
\caption{The particles and their quantum numbers under the
  $SU(3)_C \times SU(2)_L \times U(1)_Y \times U(1)_X$ and $U(1)_{\chi} $ gauge symmetry in scenario IV. Here,
  the correct $U(1)_{\chi}$ and $U(1)_X$ charges are their charges in the above table divided
  by $2{\sqrt{10}}$ and $\frac{16\sqrt{5010-900\sqrt{26}}}{67}$. }
\label{Particle-Spectrum-GSO10-6}
\end{table}

\section{Conclusion}\label{sec:summary}

We constructed the  family universal $U(1)_X$ models with three right-handed neutrinos
by choosing the $U(1)_X$ gauge symmetry as a linear combination of $U(1)_Y\times U(1)_{\chi}$ 
of $SO(10)$. To be consistent with the quantum gravity effects, 
we introduced a Dirac fermion  $\chi$ as a DM candidate,
which is odd under the gauged $Z_2$ symmetry after $U(1)_X$ breaking.
To satisfy the LUX, PANDAX, and XENON1T experimental constraints,
we found that the isospin violation DM with $f_n/f_p = -0.7$ can be realized naturally.
In addition, we studied the masses and mixings for Higgs and gauge bosons, 
considered the LHC constraints 
on the $Z'$ mass, simulated various constraints from DM direct and 
indirect detection experiments, and then presented the viable parameter spaces.
To study the LHC $Z'$ mass bounds on the generic $U(1)_X$ models, we considered four kinds of scenarios:
scenarios I, II, and III have zero  $U(1)_X$ charges, respectively, 
for quark doublets, right-handed up-type quarks,
and right-handed down-type quarks, and scenario IV has approximately equal charges for all the quarks.
We found that the LHC low bounds on the $Z'$ masses are about 4.94, 4.87, 5.34, and 5.09 TeV 
for scenarios I, II, III, and IV, respectively. Thus, scenario II 
can relax the LHC $Z'$ mass bound a little bit.


\begin{acknowledgments}

This work is supported in part by the National Key Research and Development Program of China Grant No. 2020YFC2201504, by the Projects No. 11875062, No. 11947302, No. 12047503, and No. 12275333 supported by the National Natural Science Foundation of China, by the Key Research Program of the Chinese Academy of Sciences, Grant NO. XDPB15, by the Scientific Instrument Developing Project of the Chinese Academy of Sciences, Grant No. YJKYYQ20190049, and by the International Partnership Program of Chinese Academy of Sciences for Grand Challenges, Grant No. 112311KYSB20210012.

\end{acknowledgments}


\appendix
\section{$Z'$ Decay Widths}
\label{app:width}

We present the $Z'$ decay widths in details for Sec.\ref{Particle-Spectrum-GSO10-2}.
 The vacuum value $v_s$ can be written as 
\begin{eqnarray}
v_s^2=\frac{2 M_{Z}^{4}- M_{Z}^{2} v_{h}^{2}\left(g_{2}^{2}+g_{Y}^{2}+4g_{X}^{2} X_{h}^{2}\right)}{2g_{X}^{2}\left(2 M_{Z}^{2}-\left(g_{2}^{2}+g_{Y}^{2}\right) v_{h}^{2}\right) X_{s}^{2}}~.~\,
\end{eqnarray}

The decay widths of $Z'$ to $W^+W^-$ and to $Zh$ are, respectively,
\begin{eqnarray}
&&\Gamma(Z'\rightarrow W^+W^-) 
\nonumber\\&&= \frac{g_2^2}{192\pi}\cos ^2(\theta _W) \sin ^2(\theta _{X})M_{Z'}(\frac{M_{Z'}}{M_{Z}})^4(1-4\frac{M_W^2}{M_{Z'}^2})^{3/2}(1+20\frac{M_W^2}{M_{Z'}^2}+12\frac{M_W^4}{M_{Z'}^4})~,~\,
\end{eqnarray}

\begin{eqnarray}
&&\Gamma(Z'\rightarrow Zh) 
\nonumber\\&&= \frac{g_2^2M_Z^2}{192\pi M_W^2}M_{Z'}\sqrt{\lambda}(\lambda +12\frac{M_Z^2}{M_{Z'}^2})[(\frac{4M_Z^2}{v_h^2}-g_X^2)\sin (2\theta _{X})+(\frac{4M_Z^2g_X^2}{v_h^2})\cos (2\theta _{X})]~,~\,
\end{eqnarray}
where 
\begin{eqnarray}
\lambda = 1 + (\frac{M_Z^2}{M_{Z'}^2})^2+ (\frac{M_h^2}{M_{Z'}^2})^2-2(\frac{M_Z^2}{M_{Z'}^2})-2(\frac{M_h^2}{M_{Z'}^2})-2(\frac{M_Z^2}{M_{Z'}^2})(\frac{M_h^2}{M_{Z'}^2})~.~\,
\end{eqnarray}

For our model, $\theta_{X}$ is the $Z-Z^{\prime}$ mixing angle, which is very small, so we have $\sin (\theta _{X})\approx \theta _{X}$, $\cos (\theta _{X})\approx 1$, and 
\begin{eqnarray}
\sin (\theta _{X}) =\sqrt{\frac{A-B}{2A}}~,~\,
\end{eqnarray}
where $A$ and $B$ are defined by
\begin{equation}
       \begin{split}
        A&=\sqrt{v_{h}^{4}\left(g_{2}^{2}+ g_{Y}^{2}+4 g_{X}^{2} X_{h}^{2}\right)^{2}-8 g_{X}^{2} v_{h}^{2} v_{s}^{2}\left(g_{2}^{2}+g_{Y}^{2}-4g_{X}^{2} X_{h}^{2}\right) X_{s}^{2}+16g_{X}^{4} v_{s}^{4} X_{s}^{4}}\\
        B&=4g_{X}^{2} v_{s}^{2} X_{s}^{2}-v_{h}^{2}\left(g_{2}^{2}+g_{Y}^{2}-4g_{X}^{2} X_{h}^{2}\right) ~.~\,\\
       \end{split}
\end{equation}

For DM annihilation cross sections calculation, we also need $Z'$ decay widths to $\chi \chi$ and $q\bar{q}$
    	\begin{eqnarray}
    	\Gamma_{Z'}=\Gamma(Z'\rightarrow \chi\bar{\chi})+\sum_q c_q\Gamma(Z'\rightarrow q\bar{q})~,
    	\end{eqnarray}    	
 with
    	\begin{eqnarray}
    	\Gamma(Z'\rightarrow q\bar{q})&=&\frac{m_{Z'}}{12\pi}(g_{q_A}^2\xi_q (1+\frac{2m_q^2}{m_{Z'}^2})+g_{q_V}^2\xi_q^3),\\
	\Gamma(Z'\rightarrow \chi\bar{\chi}) &=&\frac{m_{Z'}}{12\pi}g_{\chi}^2(\xi_\chi (1+\frac{2m_\chi^2}{m_{Z'}^2})+\xi_\chi^3)~.
    	\end{eqnarray}    	




\end{document}